\documentclass[useAMS,usenatbib]{mnras}

\usepackage{amsmath}
\usepackage{graphicx}
\usepackage[usenames,dvipsnames]{color}

\title{Anticipating the Sun's heavy-element abundance}
\author[D.~O.~Gough]
       {D. O. Gough\thanks{E-mail: douglas@ast.cam.ac.uk}\\
        Institute of Astronomy and Department of Applied Mathematics
              and Theoretical Physics, \\
              University of Cambridge, Cambridge CB3 0HA, UK\\
              Physics Department, Stanford University, CA 94305, USA
       }

\begin{document}



\label{firstpage}

\maketitle

\begin{abstract}
Much of our understanding of the internal structure of the Sun derives from so-called standard theoretical 
solar models.  Unfortunately, none of those models agrees completely with observation.  The discrepancy 
is commonly associated with chemical abundance, and has led to what is now called the solar abundance 
problem, the resolution of which has  previously been out of sight.  But now the Borexino Collaboration, 
who recently announced measurements of the pp-chain solar neutrinos, are optimistic that 
they will be able to measure the flux $\Phi_{\rm CNO}$ of the neutrinos emitted by the relatively weak 
CNO cycle.   Since C, N and O constitute the majority of the heavy elements, that measurement will permit 
a crucial determination of the heavy-element abundance $Z_{\rm c}$ in the Sun's energy-generating core, thereby 
shedding important light on the problem.  To accomplish that determination, a robust relation between $Z_{\rm c}$ 
and  $\Phi_{\rm CNO}$ will be required. That relation is  
$Z_{\rm c}= 0.400\, \Phi_{\rm CNO}$, where $\Phi_{\rm CNO}$ is in units of $10^{10} {\rm cm}^{-2}{\rm s}^{-1}$.

\end{abstract}

\begin{keywords}
Sun: neutrinos -- Sun: helioseismology -- Sun: abundances -- opacity
\end{keywords}

\section{Introduction}
\label{sec:introduction}
The recent report  by 
\citet{borexino2018} of the impressive 
measurement of pp-chain solar
neutrinos has provided the first observational 
confirmation of the agreement (to within about 10\%) between the nuclear energy 
production rate in the Sun and the radiant luminosity, confirming that the Sun is 
more-or-less 
in thermal balance, as it is normally assumed to be.  The result is a significant step 
towards deepening our understanding of the inner workings of a star  

The collaboration relate their results to two modern sets of standard solar models 
(SSMs) published by \citet{2017ApJ...835..202V}, one with the commonly adopted 
GS98 \citep{1998SSRv...85..161G}  abundances,
the other with the more recent, and lower, AGSS09 \citep{Asplundetal2009ARA&A}   values, 
which appear to be 
better representations of photospheric values \citep[see also][]{2011SoPh..268..255C}.   
By comparing $^{7}{\rm Be}$ and $^{8}{\rm B}$ fluxes, suitably 
adjusted for flavour transitions, with theoretical values from the models, the 
collaboration report  
a preference for the former composition, which is in conflict with the standard 
assumption that, aside from gravitational 
settling and radiative levitation,  abundances in the radiative 
envelope are the same as those in the photosphere.  However, the preference is at least 
comforting because it has long been known that solar models 
with low heavy-element abundance, $Z$,    
are ruled out by helioseismology 
\citep{DOGBridgeingap1983Natur,duvallharveybridgeingap1983Natur,dirtysunobs1998Obs.118.25C,basuantia2008PhR...457..217B}.  \citet{2017ApJ...835..202V} present a careful demonstration of this issue in the specific case of the GS98 and AGSS09 abundances. 
The conflict has been named the solar abundance problem.  

\section{Circumventing opacity}
\label{sec:circumvention}
Opacity, which depends directly on $Z$, is the principal cause of the model differences, via its control of the flow 
of heat in the Sun's radiative envelope.  A lower opacity requires a smaller temperature 
gradient, and hence a lower temperature, leading to a lower sound speed in the radiative
envelope, although the latter is offset in the energy-generating core by the higher hydrogen 
abundance, $X$, required to fuel the otherwise slower nuclear reactions. 
A reliable knowledge of opacity is therefore a crucial ingredient for understanding the 
`so-called' standard structure and evolution of the Sun.    
Its calculation involves very complicated physics, and the outcome has commonly been questioned.

It should be appreciated, however, that nuclear reactions do not themselves depend 
directly on opacity.  Moreover, the pertinent properties of the internal structure 
of the Sun can be determined, subject to relatively minor additional assumptions, by 
seismological analysis of acoustic modes of oscillation.  
Acoustic propagation depends on relatively simple physics, 
rendering correctly interpreted, yet admittedly more limited, inferences from 
helioseismology more reliable than those from SMMs.  

Setting aside an acoustic glitch that is present immediately beneath the convection zone, 
whose origin at least in part is  material redistribution in the tachocline  
\citep{elliottdogtachthickness1999ApJ...516..475E,2018MNRAS.477.3845C},  
ignored in SSMs   
but having a relatively minor impact on the overall stratification of the deep
interior of the star \citep{2017ApJ...835..202V}, 
the principal flaw in the construction of SSMs is likely to be either an error in  
the opacity calculation, a possibility that is now not wholly accepted 
(but there is, in particular, a degree of acceptance resulting from recent laboratory 
measurements by \citet{baileyetalironopacity2015Natur}), or it is the 
standard assumption that photospheric abundances directly reflect the abundances in the radiative 
envelope  \citep{guzik_mussack_2010ApJ...713.1108G}.  
A potential resolution could be, for example, that  
mechanical waves generated off-resonance at the base of the convection zone 
\citep[e.g.][]{press1981ApJ...245..286P} have amplitudes enough to carry a significant, 
yet uncertain, fraction of the total luminosity.  
If that were so, the role of opacity  would be substantially diminished.  The 
associated wave momentum flux  
would have negligible influence on the hydrostatic balance, which alone (aside from the 
adiabatic exponent $\gamma_1$, the theory of which is relatively robust in the radiative 
interior) determines the Sun's seismic structure. However, the thermal structure would then depend on 
the uncertain details of the wave spectrum.  Without knowing the thermal structure, 
no SMM can be trusted. 

\section{Determining $Z$ via seismology}
\label{sec:seismology}
As the Borexino Collaboration point out, their anticipated measurement of the neutrino flux 
$\Phi_{\rm CNO}$ produced by the CNO cycle will provide a direct evaluation of the 
CNO abundances, the dominant contributors to $Z$, irrespective of opacity.  
To accomplish that, a relation between $Z$ and $\Phi_{\rm CNO}$ will be needed.  
One might be tempted to interpolate between SMMs such as those provided by 
\citet{2017ApJ...835..202V}, but the outcome would be subject to their reliability,  
which, as I pointed out above, is in doubt.  
However, an adequate estimate of pertinent conditions in the 
core can be obtained more reliably from helioseismology, provided one adopts an assumption such as a core that was initially 
homogeneous and which suffered no material redistribution of the products of the nuclear 
reactions during the subsequent main-sequence evolution, an 
assumption that is adopted also in 
creating SMMs.  For the purpose of estimating $Z$, it is adequate to linearize 
the difference between the seismic structure of the Sun and an appropriate accurately computed SSM which,   
in the case of the representation I use here \citep{DOGLorentz2004AIPC}, was Model S of 
\citet{1996Sci...272.1286C}.  The outcome is an estimate of the sound speed $c$ and 
the density $\rho$, which are subject to an uncertainty of a few parts in $10^3$ 
\citep[cf.][]{mtdog2003ESASP.517..397T,basuantia2008PhR...457..217B},
from which the pressure $p$ can be determined from the constraint 
of hydrostatic support.  It is then necessary to estimate 
the helium abundance $Y$ and the temperature $T$, neither of which is seismically accessible. 
To this end a shell representing the tachocline in Model S was homogenized with the convection zone, and then a constant $\delta Y$ was added to $Y$, enabling $T$ to be determined 
implicitly from 
$c$, $p$ and $\rho$ with the help of the equation of state: 
$\delta Y$ and $T$ were determined simultaneously by requiring that the 
nuclear energy generation rate in the core is equal to the observed luminosity 
at the surface.  The resulting structure, Model Ss, is presented by \citet{DOGLorentz2004AIPC}.  
To the precision required here, that structure is independent of the reference SSM 
adopted \citep[cf.][]{basupinsonneaultbahcall2000ApJ...529.1084B}, and of the imperfect \citep[cf.][]{2018MNRAS.477.3845C} representation of the tachocline.

The relation between the central 
heavy-element abundance $Z_{\rm c}$ and the CNO neutrino fluxes was obtained using 
the cross-sections adopted by \citet{2017ApJ...835..202V}.  It was achieved by scaling 
model B16-GS98 of \citet{2017ApJ...835..202V} to the seismic structure of the Sun, 
in the form of Model Ss, 
using the functional forms of the CNO reaction rates, 
presumed to be in equilibrium. The relative abundances of 
C, N and O are thereby determined, irrespective of their initial values 
\citep[e.g.][]{claytonbook1983}. 
The outcome is that the Sun produces neutrino fluxes $\Phi_{13} = 1.41$, 
$\Phi_{15} = 1.06$ and $\Phi_{17} = 0.027$ per $Z_{\rm c}$, all in units of 
$10^{10} {\rm cm}^{-2}{\rm s}^{-1}$, from the beta decays of $^{13}$N,  $^{15}$O and $^{17}$F respectively.   
The central heavy-element abundance is related to the the total CNO neutrino flux according to
\begin{equation}
Z_{\rm c}= \alpha  \Phi_{\rm CNO}\,,
\label{eq:ZPhi}
\end{equation}
where $\alpha = 0.400$ and $\Phi_{\rm CNO}$ is also in units of $10^{10} {\rm cm}^{-2}{\rm s}^{-1}$. The uncertainty in $\alpha$ is dominated by uncertainties in the nuclear 
reaction rates, which are detailed by 
\citet{borexino2018} and \citet{2017ApJ...835..202V}.

Unlike hydrogen and helium abundances, the total CNO abundance is unaltered by 
the nuclear reactions, so the relation between $Z_{\rm c}$ and the heavy-element 
abundance Z throughout the radiative envelope 
is relatively secure, depending only on the weak variation resulting from 
gravitational settling.   Hence one can be assured that the analysis leading to 
equation (\ref{eq:ZPhi}) is reliable.  We now await the promised future 
measurement by the Borexino Collaboration to resolve the abundance issue.

I thank T. Sekii and the referee for useful suggestions.

\bibliographystyle{mnras}

\bibliography{refs}

\bsp

\label{lastpage}

\end{document}